\documentclass[twocolumn,showpacs,nofootinbib]{revtex4}

\input{epsf}

\def\CQG{{\it Class. Quantum Gravity} }

\def\JCAP{{\it JCAP} }

\def\NP{{\it Nucl. Phys.} }
\def\PL{{\it Phys. Lett.} }
\def\PR{{\it Phys. Rev.} }

\def\RMP{{\it Rev. Mod. Phys.} }

\def\al{\alpha} \def\be{\beta}  \def\de{\delta}
\def\ep{\epsilon}   
\def\th{\theta}   \def\ka{\kappa}

\def\om{\omega}  \def\De{\Delta} 
\def\La{\Lambda}   
 \def\Om{\Omega} \def\mn{{\mu\nu}} \def\cl{{\cal L}}

 \def\frac#1#2{{\textstyle{{#1}\over
{#2}}}} 
\def\lsim{\mathrel{\rlap{\lower4pt\hbox{\hskip1pt$\sim$}}
\raise1pt\hbox{$<$}}}
\def\gsim{\mathrel{\rlap{\lower4pt\hbox{\hskip1pt$\sim$}}
\raise1pt\hbox{$>$}}} \def\sqr#1#2{{\vcenter{\vbox{\hrule height.#2pt
\hbox{\vrule width.#2pt height#1pt \kern#1pt \vrule width.#2pt} \hrule
height.#2pt}}}}
\def\square{\mathchoice\sqr66\sqr66\sqr{2.1}3\sqr{1.5}3}
\def\beq{\begin{equation}} \def\eeq{\end{equation}}
\def\beqa{\begin{eqnarray}} \def\eeqa{\end{eqnarray}}
\def\eq#1{Eq. (\ref{#1})}

\begin{document}

\title{Modified Friedmann Equation from Nonminimally Coupled Theories of Gravity}

\author{Orfeu Bertolami\footnote{Also at Instituto de Plasmas e F\'isica Nuclear, Instituto Superior T\'ecnico, Av. Rovisco Pais 1, 1049-001, Lisboa Portugal.}}
\email{orfeu.bertolami@fc.up.pt}

\author{Jorge P\'aramos}
\email{jorge.paramos@fc.up.pt}

\affiliation{Departamento de F\'isica e Astronomia, Faculdade de Ci\^encias, Universidade do Porto, \\Rua do Campo Alegre 687, 4169-007 Porto, Portugal}

\date{\today}

\begin{abstract}

In this work we study how nonminimally coupled theories of gravity modify the usual Friedmann equation, and develop two methods to treat these. The ambiguity in the form of the Lagrangian density of a perfect fluid is emphasized, and the impact of different dominant matter species is assessed. The Cosmological Constant problem is also discussed.

\end{abstract}

\pacs{04.20.Fy, 04.80.Cc, 97.10.Cv}

\maketitle 

%%%%%%%%%%%%%%%%%%%%%%%%%%%%%%%%%%%%%%%%%%%%%%%%%%%%%%%
%%%%%%%%%%%%%%%%%%%%%%%%%%%%%%%%%%%%%%%%%%%%%%%%%%%%%%%

\section{Introduction}

Despite its great experimental success (see {\it e.g.} Refs. \cite{Will2006,OBJP2012}), it is well known that General Relativity (GR) is not the most encompassing way to couple matter with curvature. Indeed, these can be coupled, for instance, in a nonminimal way \cite{model1} (see also Refs. \cite{model2,model3,model4} for early proposals in cosmology), a fact that can have a bearing on the dark matter \cite{DM1,DM2} and dark energy \cite{DE1,DE2} problems, as well as inflation \cite{preheating,Starobinsky} and structure formation \cite{perturbations}. This putative nonminimal coupling modifies the well-known energy conditions \cite{EC} and can give rise to several implications, from Solar System \cite{PPN} and stellar dynamics \cite{solar,Martins,collapse,spherical} to close time-like curves \cite{RZF} and wormholes \cite{Lobo}.

Another interesting issue that arises in the context of gravity theories with a nonminimal coupling between curvature and matter is the fact that the Lagrangian degeneracy in the description of a perfect fluid, encountered in GR \cite{Schutz,Brown} is lifted \cite{Lagrangian}: indeed, since the Lagrangian density explicitly appears in the modified equations of motion, two Lagrangian densities leading to the same energy-momentum tensor have different dynamical implications, whereas in GR they are physically indistinguishable.

In what follows, two methods to relate modifications of the Friedmann expansion rate equation with the functions of the scalar curvature appearing in the action functional are developed --- considering  both a non-linear curvature term and a non-minimal coupling between the latter and matter. The impact of the form of the Lagrangian density in this identification is assessed, with no particular choice for this quantity or the dominant matter species (aside from the assumption of a perfect fluid). This work aims at complementing a previous study of the impact of a NMC in a cosmological context \cite{DE1}.

This work is structured as follows: in section II, we describe the nonminimally coupled (NMC) models of interest, and consider its impact in cosmology in section III; in sections IV and V, we discuss two different regimes of the NMC models; sections VI --- X concern specific explicit solutions; in section XI, we address the issue of generating a cosmological constant; finally, in section XII we present our conclusions.

%%%%%%%%%%%%%%%%%%%%%%%%%%%%%%%%%%%%%%%%%%%%%%%%%%%%%%%
%%%%%%%%%%%%%%%%%%%%%%%%%%%%%%%%%%%%%%%%%%%%%%%%%%%%%%%

\section{The model}

One considers a model that exhibits a non-minimal coupling between geometry and matter, as expressed by the action functional \cite{model1},

\beq S = \int \left[ \ka f_1(R) + f_2(R) \mathcal{L} \right] \sqrt{-g} d^4 x~~, \label{model}\eeq

\noindent where $f_i((R)$ ($i=1,2$) are arbitrary functions of the scalar curvature, $R$, $g$ is the determinant of the metric and $\ka = c^4/16\pi G$.

Variation with respect to the metric yields the modified field equations,

\beqa  \left(  F_1 + {F_2 \cl \over \ka} \right) G_\mn &=&   {1 \over 2\ka } f_2 T_\mn + \De_\mn \left(  F_1 + {F_2 \cl \over \ka} \right) + \nonumber \\  && {1 \over 2} g_\mn \left( f_1 - F_1 R - {F_2 R \cl\over \ka} \right) ~~,\label{field0} \eeqa

\noindent with $F_i \equiv df_i(R)/dR$ and $\De_\mn \equiv \nabla_\mu \nabla_\nu - g_\mn \square$. As expected, GR is recovered by setting $f_1(R) = R $ and $f_2(R) = 1$.

The trace of \eq{field0} reads

\beq \left( \ka F_1  + F_2  \cl \right) R  =   {1 \over 2} f_2 T -3 \square \left( \ka F_1 + F_2 \cl \right) + 2 \ka f_1 ~~. \label{trace0}\eeq

Resorting to the Bianchi identities, one concludes that the energy-momentum tensor of matter may not be (covariantly) conserved, since

\beq \nabla_\mu T^\mn={F_2 \over f_2}\left(g^\mn \cl-T^\mn\right)\nabla_\mu R~~, \label{cov} \eeq

\noindent can be non-vanishing. Given the analogy between the nonminimally coupled $f(R)$ theories here considered and a two-scalar-tensor theory \cite{scalar}, this can be interpreted as due to an energy exchange between the perfect fluid and the latter.

%%%%%%%%%%%%%%%%%%%%%%%%%%%%%%%%%%%%%%%%%%%%%%%%%%%%%%%
%%%%%%%%%%%%%%%%%%%%%%%%%%%%%%%%%%%%%%%%%%%%%%%%%%%%%%%

\section{Cosmology}

In order to study the effect of the modified dynamics arising from a NMC, one assumes a spatially flat, homogeneous and isotropic Universe, thus considering the Friedmann-Robertson-Walker (FRW) line element,

\beq ds^2 = - dt^2 + a^2(t) d V~~,\eeq

\noindent where $a(t)$ is the scale factor, and it is assumed that matter is a perfect fluid with energy-momentum tensor,

\beq T_\mn = (\rho + p)u_\mu u_\nu + p g_\mn \rightarrow T = 3p - \rho~~, \eeq

\noindent with $u^\mu u_\nu = -1 $ and $ u^\mu u_{\mu;\nu} = 0$; since one uses comoving coordinates, $u^\mu = \de^\mu_0$ and

\beq T_{00} = \rho ~~~~,~~~~T_{rr}={T_{\th\th}\over r^2} = a^2 p~~. \eeq

The $tt$ component of the field Eqs. (\ref{field0}) yields the modified Friedmann equation,

\beqa \label{FriedmannEq0} H^2  &= &{h(R,\cl,\rho) \over 6 \ka}~~, \\ \nonumber  h(R,\cl,\rho) &\equiv&  {\ka  \over F_1 + {F_2 \cl \over \ka} }\bigg[ {f_2 \rho \over \ka} - 6 H\partial_t \left(  F_1 + {F_2 \cl \over \ka} \right) + \\ \nonumber && \left(  F_1 + {F_2 \cl\over \ka}\right) R - f_1  \bigg] ~~, \eeqa

\noindent where $H \equiv \dot{a}/a$; clearly, for $f_1(R) = R $ and $f_2(R) = 1$, the usual result $h(\rho) = \rho$ ensues.

The purpose of this work is to ascertain the forms for $f_1(R) $ and $f_2(R)$ compatible with a specific form $h(\rho)$ for the Friedmann-like equation; this requires that both the Lagrangian density $\cl$ and the scalar curvature $R$ are expressed in terms of the energy density $\rho$. The former is simply attained by writing

\beq \label{choice} \cl = -\al \rho~~~~,~~~~~\al = \cases{ 1~~~~,~~~~\cl = -\rho  \cr -\om ~~~~,~~~~\cl = p } ~~,\eeq

\noindent where $\om = p/\rho$ is the equation of state (EOS) parameter. This captures the two possible Lagrangian formulations of a perfect fluid (although it was argued in Ref. \cite{Lagrangian} that $\cl = -\rho$ is the most adequate choice), including the possibility of effectively describing a scalar field where $\cl = p(\phi, \dot{\phi})$.

Replacing \eq{choice} into \eq{FriedmannEq0}, one gets 

\beqa \label{FriedmannEq1} h(R,\rho) &\equiv& {\ka  \over F_1 - \al F_2 {\rho \over \ka} }\bigg[ {f_2 \rho \over \ka} - f_1 + \\ \nonumber && 6 H\partial_t \left(  F_1 -\al F_2 {\rho \over \ka} \right) + \left(  F_1 -\al F_2 {\rho \over \ka}\right) R \bigg]~~, \eeqa

\noindent while the trace \eq{trace0} yields

\beqa  \label{trace1}&& \left( F_1  - \al {F_2  \rho \over \ka} \right) R  = \\ \nonumber && -{1 - 3\om \over 2} {f_2 \rho \over \ka} -3 \square \left( F_1 -\al {F_2 \rho \over \ka} \right) + 2 f_1 ~~,\eeqa

\noindent and the non-trivial $\nu = t$ component of the energy conservation \eq{cov} becomes

\beq \dot{\rho} + 3H(1+\om) \rho ={F_2 \over f_2}(\al -1) \rho \dot{R}~~. \label{cov1} \eeq

In the following sections one derives the aforementioned relation between the functions $f_1(R)$, $f_2(R)$ and $h(\rho)$, using two different methods: firstly, one considers only a perturbative regime, so that the scalar curvature is approximately given by its usual GR expression, $R(\rho) = R_0(\rho) \equiv -T/(2\ka)$. Secondly, one attempts to obtain a correspondence valid even in a non-perturbative scenario, by imposing instead that the solution to the above system of equations obeys the condition $\ka F_1 + F_2 \cl = {\rm constant}$.

%%%%%%%%%%%%%%%%%%%%%%%%%%%%%%%%%%%%%%%%%%%%%%%%%%%%%%%
%%%%%%%%%%%%%%%%%%%%%%%%%%%%%%%%%%%%%%%%%%%%%%%%%%%%%%%

\section{Perturbative regime}

In this section, one establishes a relation between the non-trivial forms for $f_1(R)$, $f_2(R)$ and the modified Friedmann equation. Following Ref. \cite{Sotiriou}, one writes

\beq \label{perturbative} f_1(R) = R + \phi_1(R)~~, ~~f_2(R) = 1 + \phi_2(R)~~, \eeq

\noindent where $\phi_i(R)$ are assumed to be perturbative, $\phi_1(R) \ll R$ and $\phi_2(R) \ll 1$. By the same token, the scalar curvature is assumed to be approximately given by

\beq R(\rho) \approx R_0(\rho) = (1-3\om) {\rho \over 2\ka } ~~, \eeq

\noindent so that

\beq {d\phi_i \over dR} = {2\ka \over 1-3\om} \phi_i'(\rho)~~, \eeq

\noindent where the prime denotes differentiation with respect to the energy density $\rho$. This perturbative approach cannot be applied for radiation or relativistic matter, as both have an EOS parameter $\om =1/3$, so that the ensuing curvature vanishes according to GR, $R_0=0$.
 
Using  the covariant conservation of the energy-momentum tensor, \eq{cov1} and writing $h(\rho) = \rho + \de(\rho) $, \eq{FriedmannEq1} becomes, to first order in $\phi_i$,

\beqa \label{FriedmannEqPert1} \de(\rho) &=& - \ka \phi_1 + \left( \phi_2 -\ka {1+3\om \over 1-3\om}\phi_1' \right) \rho + \\ \nonumber && { 6 (1+\om)\ka \phi''_1 - \al (5+3\om ) \phi_2' \over 1-3\om} \rho^2 - \\ \nonumber && {6 (1+\om) \al \phi''_2 \over 1-3\om} \rho^3 ~~. \eeqa

This linear non-homogeneous differential equation enables a direct translation between the form of the modified Friedmann equation and the non-trivial forms for $f_1(R)$, $f_2(R)$ giving origin to it. Setting $\phi_i(R)=0$ trivially yields $\de(\rho) =0$, while the inclusion of a Cosmological Constant (CC) term in the action, $\phi_1(R)= -2 \La$ and $\phi_2(R)=0$, leads to $ H^2 = (\rho/6\ka) + \La / 3 $, as expected. Notice, however, that a dominant CC, which occurs since $z \sim 0.4$ \cite{Gong}, cannot be accommodated by the mechanism outlined above, as it is non-perturbative, {\it i.e.} $R \sim 4\La \neq R_0$ (since $\Om_\La \sim 0.7 > \Om_m \sim 0.3$).

\subsection{``Neutral'' solutions}

One notices that, since \eq{FriedmannEqPert1} is linear, its general solution for a given modification $\de(\rho)$ of the Friedmann equation includes the solution of the corresponding homogeneous differential equation, {\it i.e.}, the one for which $\de(\rho)=0$. Physically, this simply states that any perturbation $\phi_{1H}(R)$ in the action has a counterpart $\phi_{2H}(R)$ that cancels out its dynamical effect. These functions, here dubbed as ``neutral'', are related by 

\beqa \label{FriedmannEqPert1homo1} && 0 = - \ka \phi_{1H} + \left( \phi_{2H} -\ka {1+3\om \over 1-3\om}\phi_{1H}' \right) \rho + \\ \nonumber && { 6 (1+\om)\ka \phi''_{1H} - \al (5+3\om ) \phi_{2H}' \over 1-3\om} \rho^2 - {6 (1+\om) \al \phi''_{2H} \over 1-3\om} \rho^3  ~~. \eeqa

By the same token, if one sets $\phi_{2H} = 0$ and solves the above differential equation for $\phi_{1H}$ (and vice-versa), one finds that functions of the form $ \phi_{1H}(R) \sim  R^{n_1} $ and $\phi_{2H}(R) \sim R^{n_2}$, with

\beqa \label{excludedExp} n_1 &=&{7 + 9 \om \pm \sqrt{ 73 + 78 \om + 9 \om^2} \over 12 (1 + \om)} ~~, \\ \nonumber n_2 &=& { 1+3\om \pm \sqrt{ (1+3\om)^2 + {24 \over \al} (1+\om)(1-3\om) } \over 12 (1 + \om)} ~~,\eeqa

\noindent do not modify the Friedmann equation at first order, since their contribution to \eq{FriedmannEqPert1} vanish.

For completeness, one lists the values of the above exponents for the relevant matter species:

\begin{itemize}
	\label{excludelist}
	
	\item Non-relativistic dust ($\om=0$, $\al = 1$):
		\beq n_1= {7 \pm \sqrt{ 73 } \over 12 } ~~~~,~~~~ n_2 = { 1 \pm 5 \over 12}~~.\eeq
		
	\item Ultrastiff matter ($\om=1$, $\al = 1$):
		\beq n_1 = { 4 \pm  \sqrt{10} \over 6}~~~~,~~~~n_2 = {1 \pm i \sqrt{5} \over 6} ~~.\eeq

	\item Scalar field ($\om=1$, $\al = -\om = -1$):
		\beq n_1 = { 4 \pm  \sqrt{10} \over 6}~~~~,~~~~n_2 = {1 \pm \sqrt{7} \over 6} ~~.\eeq
		
\end{itemize}

Notice the unphysical, complex value of $n_2$ for ultrastiff matter. The conditions under which the method outlined here can be applied to a power-law modification of the Friedmann equation will be discussed in a subsequent section.

%%%%%%%%%%%%%%%%%%%%%%%%%%%%%%%%%%%%%%%%%%%%%%%%%%%%%%%
%%%%%%%%%%%%%%%%%%%%%%%%%%%%%%%%%%%%%%%%%%%%%%%%%%%%%%%

\section{Non-perturbative, relaxed regime}
\label{relaxedregime}

In this section one obtains a correspondence between the functions $f_1(R)$, $f_2(R)$ and the {\it r.h.s.} of the Friedmann \eq{FriedmannEq0}; instead of assuming a perturbative regime, the possibility that the scalar curvature may strongly deviate from its GR expression, $R(\rho) \neq R_0(\rho)$ is addressed.

To tackle this situation, the hypothesis that the combination $ F_1 + F_2 \cl/\ka $ is constant is posited. This could stem, for instance, from a fixed point in the dynamical system constituted by \eq{field0}: indeed, the presence of the kinetic term $\De_\mn (F_1 + F_2 \cl/\ka)$ in these is suggestive that some fixed points could obey the constraint

\beq \label{constraint} F_1 + {F_2 \cl \over \ka} = F_1 - \al F_2 {\rho \over \ka} = A \neq 0 ~~.\eeq

\noindent Notice that GR yields $A=1$, so that one expects this quantity to be either always equal to unit, or to depend on additional parameters related to the functions $f_1(R)$, $f_2(R)$ in such a way that $A\rightarrow 1$ renders GR.
 
Inserting \eq{constraint} into \eq{FriedmannEq1} yields

\beqa \label{FriedmannEq2} Ah = f_2 \rho + A \ka R - \ka f_1 ~~, \eeqa

\noindent while the trace \eq{trace1} becomes

\beq A R  = -{1 - 3\om \over 2\ka } f_2 \rho + 2 f_1 ~~. \label{trace2}\eeq

Solving for $f_1(R)$ and inserting back into \eq{FriedmannEq2}, one gets

\beqa \label{FriedmannEq3} 2h = \ka R + {3\over 2A } (1+\om) f_2 \rho ~~. \eeqa

The above is further simplified by writing the scalar curvature explicitly,

\beqa \label{Rexp1} R &=& 6 (\dot{H} + 2 H^2) = 6 \left( {\dot{(H^2)} \over 2H} + 2 H^2 \right) = \\ \nonumber &=&  {1 \over \ka} \left( {\dot{h}\over 2H} +2 h\right) =  {1 \over \ka} \left( { h'(\rho) \over 2}{\dot{\rho}\over H} +2 h\right) ~~.\eeqa

One may resort to \eq{cov1} to eliminate $\dot{\rho}$; for this, one assumes that the scalar curvature can be written as a function of the energy density alone, $R = R (\rho)$ (as will be shown below), so that 

\beq f_2(R) = f_2(R(\rho))\rightarrow  F_2\dot{R} = f'_2 (\rho) \dot{\rho} ~~, \eeq

\noindent {\it i.e.} one differentiates with respect to the energy density $\rho$, instead of the scalar curvature; in the remainder of this work, the notation $F_i = df_i(R)/dR$ and $f'_i = df(R(\rho))/d\rho$ is adopted, for brevity. Thus, \eq{cov1} implies that

\beq \label{cov2} \dot{\rho} = -  {3H(1+\om) f_2 \rho \over f_2 + (1 - \al )f_2' \rho }~~. \eeq

\noindent Replacing this into \eq{Rexp1} yields

\beq \label{Rexp2} R(\rho) =  {1 \over \ka} \left(2 h - {3 \over 2} {(1+\om) f_2 h' \rho \over f_2 + (1 - \al )f_2' \rho }  \right) ~~,\eeq

\noindent thus showing that the scalar curvature can be expressed solely as a function of the energy density, as claimed above.

Inserting this into \eq{FriedmannEq3} leads to 

\beq \label{FriedmannEq4} A {(1+\om) f_2 h' \rho \over f_2 + (1 - \al )f_2' \rho } = (1+\om) f_2 \rho ~~.\eeq

Assuming that $ f_2 + (1 - \al )f_2' \rho \neq 0$ and $(1+\om) f_2 \rho \neq 0$ (so that a $\om \neq -1$ scalar field in slow-roll is excluded), one finally obtains 

\beq \label{FriedmannEq5}  f_2 + (1 - \al )f_2' \rho = A h' ~~.\eeq

\noindent This relationship provides the connection between the modified form of the Friedmann equation and the NMC; it is worth remarking that, if $\cl = -\rho \rightarrow \al=1$, then the NMC is directly read, $f_2(\rho)=Ah'$. Inserting this into \eq{Rexp2}, one gets

\beq \label{Rexp3} R(\rho) =  {1 \over \ka} \left[2 h - {3 \over 2A} (1+\om) f_2 \rho \right] ~~.\eeq

Finally, \eq{trace1} allows one to read

\beq \label{f1exp1} f_1 = {1 \over \ka} \left( A h - {1+3\om \over 2} f_2 \rho \right) ~~. \eeq

In order to identify the functions $f_1(R)$ and $f_2(R)$ that enable a particular form for $h(\rho)$, one must first solve \eq{FriedmannEq5} for $f_2(\rho)$; one can then compute $R(\rho)$ from \eq{Rexp3} and invert it to obtain $\rho = \rho(R)$; replacing this back into $f_2(\rho)$ yields the aimed NMC. A similar procedure using \eq{f1exp1} yields the curvature term $f_1(R)$. Condition \eq{constraint}  can then be explicitly checked (it can also be shown to be valid for a general $h(\rho)$, although that is not shown here).

In the following sections some specific forms for the {\it r.h.s.} of the Friedmann equation, {\it i.e.} the function $h(\rho)$, are explored, with the aim of obtaining the functions $f_i(R)$ that give rise to the latter using the two methods proposed above. 

%%%%%%%%%%%%%%%%%%%%%%%%%%%%%%%%%%%%%%%%%%%%%%%%%%%%%%%
%%%%%%%%%%%%%%%%%%%%%%%%%%%%%%%%%%%%%%%%%%%%%%%%%%%%%%%

\section{Application: Cosmological Constant}
\label{CCapp}

One remarks that \eq{FriedmannEq5} is linear, so that its solution is the sum of a particular solution plus the general solution of the corresponding homogeneous equation. The latter amounts to $h'(\rho) = 0$, yielding a constant Hubble parameter $H$ ({\it i.e.} a De Sitter phase) --- and can thus be related to the CC.

Assuming that the matter contribution is negligible, one thus has $h(\rho ) \approx 6\ka H_0^2$: for this reason, one cannot rely on the perturbative approach, as the scalar curvature is approximately constant $ R = 2h / \ka = 12H_0^2$ and cannot be considered to be a small deviation from the form $R = R_0 = (1-3\om)\rho/2\ka$, as argued above. For this reason, one must rely solely on the method developed for the relaxed regime: in particular, \eq{FriedmannEq5} becomes homogeneous,

\beqa f_2 + (1 - \al )f_2' \rho = 0 \rightarrow \\ \nonumber f_2(\rho) \propto \rho^{1/(\al-1)} ~~. \eeqa

\noindent Inserting into \eq{Rexp2}, one gets a constant, as expected; as a result, one cannot invert it so to write $\rho=\rho(R)$, and the obtained NMC cannot be expressed as a function of the scalar curvature.

Thus, one concludes instead that, although the energy density may be subdominant with respect to the contribution of the CC, it must also be accounted for. This translates into the choice 

\beq \label{CC} h(\rho ) = \rho + 2\ka \La ~~, \eeq

\noindent so that \eq{FriedmannEq5} reads

\beqa \label{FriedmannEqCC0}  f_2 + (1 - \al )f_2' \rho &=& A= 1 \rightarrow \\ \nonumber  f_2(\rho ) &=& 1 + C \left({\rho \over 2\ka\La}\right)^{1/(\al-1)} ~~,\eeqa

\noindent where $C$ is an integration constant and $A=1$ was set to enforce $f(0) = 1$. If $\cl = -\rho \rightarrow \al = 1$, the exponent on the {\it r.h.s.} is singular, so one has to separately consider both options $\al = 1$ or $\al = -\om$.

%%%%%%%%%%%%%%%%%%%%%%%%%%%%%%%%%%%%%%%%%%%%%%%%%%%%%%%
%%%%%%%%%%%%%%%%%%%%%%%%%%%%%%%%%%%%%%%%%%%%%%%%%%%%%%%

\subsection{$\cl = -\rho$ case}

If $\al = 1$, then \eq{FriedmannEq5} straightforwardly reads $f_2(R)= 1$, so that a minimal coupling between curvature and matter is recovered. Naturally, \eq{Rexp3} collapses into the GR result

\beq R(\rho) =  4\La + {(1-3\om)\rho \over 2\ka} ~~,\eeq

\noindent and \eq{f1exp1} reads

\beq f_1 = 2\La +  {(1-3\om)\rho  \over 2\ka } = R - 2\La  ~~, \eeq

\noindent thus yielding the trivial result that, in the absence of a NMC, a CC can be produced by simply inserting it into $f_1(R)$.

%%%%%%%%%%%%%%%%%%%%%%%%%%%%%%%%%%%%%%%%%%%%%%%%%%%%%%%
%%%%%%%%%%%%%%%%%%%%%%%%%%%%%%%%%%%%%%%%%%%%%%%%%%%%%%%

\subsection{$\cl =-p$ case}

One is thus left with the possibility $\cl = p \rightarrow \al = -\om$, which is read directly from \eq{FriedmannEqCC0},

\beq \label{f2expCC} f_2(\rho) = 1 + C \left({ 2\ka\La \over \rho }\right)^{1/(1+\om)} ~~. \eeq

Inserting this into \eq{Rexp3} yields

\beq \label{RexpCC} R(\rho) =  4 \La +  \left[ 1 - 3\om - 3 (1+\om) C \left({2\ka\La \over \rho }\right)^{1/(1+\om)}  \right] { \rho\over 2\ka} ~~.\eeq

Although the above cannot be easily inverted in order to obtain $\rho = \rho(R)$, \eq{trace1} allows one to read

\beqa \label{f1expCC} \nonumber f_1(R) &=&  2 \La + \left[ 1-3\om  - (1+3\om ) C \left({ 2\ka\La \over \rho }\right)^{1/(1+\om)}\right] {\rho \over 2\ka} \\  &= & R - 2 \La ~~,\eeqa

\noindent so that the usual expression for GR with a CC is recovered.

One also obtains

\beq R'(\rho ) = {1 - 3\om \over 2\ka} - {3 \om C \over 2\ka } \left({ 2\ka\La \over \rho }\right)^{1/(1+\om)} ~~.\eeq

Up to now, the integration constant $C$ has not been specified. One is only left with the relaxation condition \eq{constraint} to look for its value; it reads

\beqa && F_1 + \om {F_2 \rho\over \ka} = 1 + \om{ f_2'(\rho) \rho \over \ka R'(\rho)} = \\ \nonumber && 1 - 2 {\om \over 1+\om}C { \left({ 2\ka\La \over \rho }\right)^{1/(1+\om)} \over  1 - 3\om- 3 \om C \left({ 2\ka\La \over \rho }\right)^{1/(1+\om)} } ~~, \eeqa

\noindent which is only constant (and equal to $A = 1$) if $C=0$. This, however, collapses Eqs. (\ref{f2expCC}) and (\ref{RexpCC}) to their GR form --- and so one concludes that the only way of obtaining a CC obeying the relaxation condition \eq{constraint} is by trivially setting $f_1(R) = R - 2\La$ and $f_2(R) = 1$. 

This results complements a similar argument, discussed in Ref. \cite{DE1}, which was valid only for a pure De Sitter expansion, {\it i.e.} disregarding the contribution of the energy density of matter, so that $h(\rho ) \sim 2\ka \La$: if the scalar curvature $R$ does not evolve with time, neither do $F_1$ and $F_2$; since the energy density $\rho$ varies (albeit in a modified fashion, as given by \eq{cov2}), then it is impossible to enforce the relaxation condition $F_1 - \al F_2 \rho/\ka$ (except for the trivial case obtained). Here, it is shown that this result does not change even if the matter contribution is considered. This does not preclude the possibility of obtaining a CC from a NMC, although the two methods developed above cannot be applied. Such an issue is deferred to section \ref{CCregime}.

Given the above, one drops the homogeneous solution $\sim \rho^{-1/(1+\om)}$ from the scenarios studied in the subsequent paragraphs, thus considering that the Friedmann equation (as given by $h(\rho)$) cannot include a constant term.

%%%%%%%%%%%%%%%%%%%%%%%%%%%%%%%%%%%%%%%%%%%%%%%%%%%%%%%
%%%%%%%%%%%%%%%%%%%%%%%%%%%%%%%%%%%%%%%%%%%%%%%%%%%%%%%

\section{Application: Power-law form}

In this section one addresses the possibility that the modified Friedmann equation is written as

\beq \label{Power} h(\rho) = \rho \left[ 1 + \ep \left({\rho \over \rho_c}\right)^{n-1} \right]~~~~, ~~~~n \neq 1~~. \eeq

\noindent where $\ep = \pm 1$ marks the sign of the power-law addition. Positive quadratic corrections, $\ep=1$ and $n=2$, arise in braneworld scenarios \cite{Brane1,Brane2,Brane3}, while Loop Quantum Gravity \cite{LQC} gives rise to a negative quadratic term, $\ep = -1$. The so-called Cardassian models postulate deviations of the form above \cite{Cardassian}, with $\ep =1$ and $n < 2/3$. By following the two methods here devised, one may ascertain the related form of the  functions $f_1(R)$ and $f_2(R)$.

%%%%%%%%%%%%%%%%%%%%%%%%%%%%%%%%%%%%%%%%%%%%%%%%%%%%%%%
%%%%%%%%%%%%%%%%%%%%%%%%%%%%%%%%%%%%%%%%%%%%%%%%%%%%%%%

\subsection{Perturbative regime}
\label{FriedmannEqPertPowerSection}

The perturbation $\de(\rho) \equiv h(\rho)-\rho = \ep \rho_c( \rho/\rho_c)^n $ is now considered. Inspection of  \eq{FriedmannEqPert1} shows that the functions $\phi_i(\rho)$ should also be power-laws; writing 

\beq \phi_1(\rho) = K_1 {\rho_c \over \ka} \left({\rho \over \rho_c} \right)^{n_1} ~~~~,~~~~  \phi_2(\rho) = K_2 \left({\rho \over \rho_c} \right)^{n_2}, \eeq

\noindent where $K_1$ and $K_2$ are dimensionless constants, and thus

\beqa \label{FriedmannEqPert1Power} && \ep =  - \left( 1 + { 7 + 9\om - 6 (1+\om) n_1 \over 1-3\om} n_1 \right)  K_1 \left({\rho \over \rho_c} \right)^{n_1-n} \\ \nonumber && + \left( 1 + { 1+ 3\om - 6 (1+\om) n_2 \over 1-3\om} \al n_2 \right) K_2 \left({\rho \over \rho_c} \right)^{n_2-n+1} ~~, \eeqa

\noindent so that one must have $n = n_1 = n_2+1$. Replacing into the above, one obtains the following relation between the constants $K_1$ and $K_2$:

\beqa \label{FriedmannEqPert2Power} \ep & =& - \left( 1 + { 7 + 9\om - 6 (1+\om) n \over 1-3\om} n \right)  K_1  \\ \nonumber && + \left( 1 + { 7+ 9\om - 6 (1+\om) n \over 1-3\om} \al (n-1) \right) K_2 ~~. \eeqa

The result above plainly shows that, at a perturbative level, a $\de(\rho) \sim \rho^n$ power-law modification of the Friedmann equation can be achieved by either a non-linear curvature term,

\beq \label{FriedmannEqPert2Powerf1} f_1(R) = R \left[ 1 + {2 K_1 \over 1-3\om} \left( {2 \ka R \over \rho_c(1-3\om)}\right)^{n-1} \right]~~, \eeq

\noindent a power-law NMC,

\beq \label{FriedmannEqPert2Powerf2} f_2(R) = 1 + K_2 \left({2\ka R \over \rho_c (1-3\om)} \right)^{n-1} ~~, \eeq

\noindent or a combination of both.

Furthermore, due to the linearity of \eq{FriedmannEqPert1}, one concludes that if the perturbative modification of the Friedmann equation can be expanded in powers of $\rho$,

\beq \de(\rho) = -\sum_{n=1}^\infty a_n \left( {\rho \over \rho_c}\right)^n ~~, \eeq

\noindent it is related to the functions

\beqa f_1(R) &=& R \left[ 1 + \sum_{n=1}^\infty {2 a_n K_{1n} \over 1-3\om} \left( {2 \ka R \over \rho_c(1-3\om)}\right)^{n-1} \right] ~~, \nonumber \\ f_2(R) &=& 1 + \sum_{n=1}^\infty a_n K_{2n} \left({2\ka R \over \rho_c (1-3\om)} \right)^{n-1} ~~,\eeqa

\noindent where the $K_{in}$ ($i=1,2$) coefficients obey \eq{FriedmannEqPert2Power} for each $n$.

Finally, one recalls that, from the previous discussion about homogeneous solutions of \eq{FriedmannEqPert1}, power-law perturbations $\phi_i(R)$ with exponents given by \eq{excludedExp} do not have an impact on the Friedmann equation (at first perturbative order).

This means that any power-law perturbation $\de(\rho) \sim \rho^n$ to the Friedmann equation can be considered. As an example, suppose one is aiming at implementing a $h(\rho) \sim \rho^{3/2}$ perturbation for a dust dominated universe: although a NMC $\phi_2(R) \sim R^{1/4}$ has no impact on \eq{FriedmannEqPert1Power}, the function $\phi_1 \sim R^{3/2}$ gives rise to the desired power-law modification. Conversely, $h(\rho) \sim \rho^{7+\sqrt{73} \over 12} $ requires a NMC $\phi_2(R) \sim \rho^{-5+\sqrt{73} \over 12 }$, since $\phi_1(R) \sim R^{7+\sqrt{73} \over 12} $ would have no dynamical impact in a Universe dominated by a non-relativistic perfect fluid. 

Furthermore, one notices that the listed neutral exponents are always non-integer, so they do not contradict the assumed Taylor expansion, as discussed above. 

%%%%%%%%%%%%%%%%%%%%%%%%%%%%%%%%%%%%%%%%%%%%%%%%%%%%%%%
%%%%%%%%%%%%%%%%%%%%%%%%%%%%%%%%%%%%%%%%%%%%%%%%%%%%%%%

\subsection{Relaxed regime}

One now uses \eq{FriedmannEq5} to read the NMC,

\beqa \label{FriedmannEqPower}  f_2 + (1 - \al )f_2' \rho &=& A \left [ 1 + \ep n \left({\rho \over \rho_c}\right)^{n-1} \right] \rightarrow \\ \nonumber f_2(R) &=& 1 + { \ep n \over n + \al (1-n) } \left({\rho \over \rho_c}\right)^{n-1}    ~~,\eeqa

\noindent where $A=1$ was fixed so that $f_2(\rho) = 1$ when $\rho \ll \rho_c $ and GR is recovered.
 
One may now replace the obtained expression into \eq{Rexp3}, leading to

\beq \label{RexpPower} R(\rho) =  {\rho \over 2\ka} \left[ 1 -3 \om + \ep {4 \al(1-n) + n (1 - 3 \om) \over n + \al (1-n) }\left({\rho \over \rho_c}\right)^{n-1} \right] ~~,\eeq

\noindent and, from \eq{f1exp1},

\beq \label{f1expPower} f_1 =  {\rho \over 2\ka} \left[ 1 -3 \om  +\ep { 2 \al (1-n) + n (1 - 3\om) \over n + \al (1-n) } \left({\rho \over \rho_c}\right)^{n-1} \right] ~~. \eeq

In order to write $f_1$ and $f_2$ in terms of the scalar curvature, one must invert \eq{RexpPower}, which requires solving a $n$-order algebraic equation. For a generic value of the exponent $n$ and the EOS parameter $\om$, one can proceed numerically; for completion, the examples considered in the previous section are worked out below:

\begin{itemize}
	
	\item Non-relativistic dust ($\om=0$, $\al = 1$):
		\beqa R(\rho) &=& {\rho \over 2\ka} \left[ 1 + \ep(4-3n) \left({\rho \over \rho_c}\right)^{n-1} \right] ~~, \\ \nonumber f_1(R) &=&  {\rho \over 2\ka} \left[ 1 +\ep (2-n) \left({\rho \over \rho_c}\right)^{n-1} \right] ~~, \\ \nonumber f_2(R) &=& 1 +\ep n \left({\rho \over \rho_c}\right)^{n-1} ~~.\eeqa
	
	\item Radiation ($\om=1/3$, $\al = -\om=-1/3$ and $n \neq 1/4$):
		\beqa R(\rho) &=& 2\ep {\rho_c \over \ka} { n-1 \over 4n-1}\left({\rho \over \rho_c}\right)^n ~~, \\ \nonumber f_1(R) &=&  \ep{\rho_c \over \ka} { n-1 \over 4n-1 } \left({\rho \over \rho_c}\right)^n = {R\over 2}  ~~, \\ \nonumber f_2(R) &=&  1 + \ep {3n \over  4n-1} \left({\rho \over \rho_c}\right)^{n-1}  =  \\ \nonumber &&  1 + {3n \over \sqrt[n]{ \ep(4n-1)}} \left( { \ka R \over 2(n-1)\rho_c }\right)^{(n-1)/n} ~~.\eeqa
	
	\item Relativistic matter ($\om=1/3$, $\al = 1$):
	\beqa R(\rho) &=&  2\ep (1-n){\rho_c\over \ka} \left({\rho \over \rho_c}\right)^n ~~, \\ \nonumber f_1(R) &=& 	\ep (1-n){\rho_c\over \ka} \left({\rho \over \rho_c}\right)^n= {R \over 2} ~~, \\ \nonumber f_2(R) &=&  1 +\ep n \left({\rho \over \rho_c}\right)^{n-1} = \\ \nonumber && 1 + \ep^{1/n} n \left( { \ka R \over 2(1-n)\rho_c} \right)^{(n-1)/n}   ~~.\eeqa
	
	\item Ultrastiff matter ($\om=1$, $\al = 1$):
		\beqa R(\rho) &=& - {\rho \over \ka} \left[ 1 + \ep (3n-2) \left({\rho \over \rho_c}\right)^{n-1} \right]  ~~, \\ \nonumber f_1(\rho) &=& - {\rho \over \ka} \left[ 1  + \ep (2n-1 ) \left({\rho \over \rho_c}\right)^{n-1} \right] ~~, \\ \nonumber f_2(\rho) &=&  1 + \ep n \left({\rho \over \rho_c}\right)^{n-1}    ~~.\eeqa

	\item Scalar field ($\om=1$, $\al = -\om = -1$):
		\beqa \label{scalarfieldpowerlaw} R(\rho) &=& - {\rho \over \ka} \left[ 1 + \ep{n-2 \over 2n-1}\left({\rho \over \rho_c}\right)^{n-1} \right]  ~~, \\ \nonumber f_1(\rho) &=& - {\rho \over \ka} \left[ 1  + { \ep \over 2n-1} \left({\rho \over \rho_c}\right)^{n-1} \right] ~~, \\ \nonumber f_2(\rho) &=&  1 + \ep{n \over 2n-1} \left({\rho \over \rho_c}\right)^{n-1}    ~~.\eeqa

\end{itemize}

One concludes that radiation or relativistic matter are not suited for implementing a power-law modification of the Friedmann equation, as they require that $f_1(R) = R/2$.

%%%%%%%%%%%%%%%%%%%%%%%%%%%%%%%%%%%%%%%%%%%%%%%%%%%%%%%
%%%%%%%%%%%%%%%%%%%%%%%%%%%%%%%%%%%%%%%%%%%%%%%%%%%%%%%

\section{Application: Quadratic Form}
\label{quadraticsection}

The specific case

\beq \label{LQC} h(\rho) = \rho \left( 1 +\ep {\rho \over \rho_c} \right) ~~, \eeq

\noindent is of particular interest, as it arises from braneworld scenarios \cite{Brane1,Brane2,Brane3}, Loop Quantum Gravity (where it is found to be valid even for very high densities, $\rho \lesssim \rho_c$) \cite{LQC} and, phenomenologically, it can be viewed as a leading order correction to the linear form of the Friedmann equation. One now replaces $n=2$ into the previous results, for illustration.

%%%%%%%%%%%%%%%%%%%%%%%%%%%%%%%%%%%%%%%%%%%%%%%%%%%%%%%
%%%%%%%%%%%%%%%%%%%%%%%%%%%%%%%%%%%%%%%%%%%%%%%%%%%%%%%

\subsection{Perturbative regime}

Using Eqs. (\ref{FriedmannEqPert2Powerf1}) and (\ref{FriedmannEqPert2Powerf2}) one finds that, at perturbative level, the modified Friedmann \eq{LQC} stems from both a quadratic curvature term

\beq f_1(R) = R + {4\ka K_1 \over (1-3\om)^2} {R^2 \over \rho_c}~~, \eeq

\noindent and a linear NMC 

\beq f_2(R) = 1 + {2\ka K_2 \over (1-3\om)} {R \over \rho_c } ~~, \eeq

\noindent with the constraint

\beq \label{FriedmannEqPert2PowerLQC} \ep = 9{1 +\om \over 1-3\om} K_1 - \left( { 5+ 3\om \over 1-3\om} \al - 1 \right) K_2 ~~. \eeq

\noindent The above agrees with the particular case previously studied in Ref. \cite{Sotiriou}, with a minimally coupled scalar field ($\om=-\al =1$) and a negative quadratic term, $\ep = -1$,

\beq f_1(R) = R + {\ka R^2 \over 9 \rho_c}~~~~,~~~~f_2(R) =1~~. \eeq

\noindent For comparison, one finds that a scalar field can also give rise to the same quadratic modification resorting only to a linear NMC,

\beq f_1(R)=R~~~~,~~~~f_2(R) = 1 + {R \over 3\rho_c } ~~. \eeq

\noindent Notice that, although the NMC has a positive slope, the negative curvature $R \approx -\rho/\ka < 0 $ found in \eq{scalarfieldpowerlaw} leads to a subtractive quadratic term $h(\rho ) = - \rho^2 /\rho_c$.

%%%%%%%%%%%%%%%%%%%%%%%%%%%%%%%%%%%%%%%%%%%%%%%%%%%%%%%
%%%%%%%%%%%%%%%%%%%%%%%%%%%%%%%%%%%%%%%%%%%%%%%%%%%%%%%

\subsection{Relaxed regime}

From Eqs. (\ref{FriedmannEqPower})--(\ref{f1expPower}), one gathers that

\beqa \label{ResultsLQC} && R(\rho) =  {\rho \over \ka} \left( {1 - 3\om \over 2} +  \ep{ 1-3\om - 2 \al \over 2-\al} {\rho \over \rho_c} \right) \rightarrow \\ \nonumber && \rho(R) = \ep {(2-\al)(1-3\om) \over 4(1-2\al -3\om)} \rho_c \times \\ \nonumber && \left( \sqrt{1 + \ep {16 (1-3\om -2\al ) \over (2-\al)(1-3\om)^2} {\ka R \over \rho_c} } - 1 \right) ~~, \\ \nonumber && f_1(R) = {\rho \over \ka} \left( {1-3\om \over 2} + \ep {1-3\om-\al \over 2-\al} {\rho \over \rho_c} \right) = \\ \nonumber && {1-3\om-\al \over 1- 3\om - 2\al} R - \ep { \rho_c \over 8\ka } (2-\al)\al \left({1-3\om  \over 1-3\om-2\al}\right)^2 \times \\ \nonumber &&  \left( \sqrt{1 + \ep {16 (1-2\al -3\om) \over (2-\al)(1-3\om)^2} {\ka R \over \rho_c} } - 1 \right)   ~~, \\ \nonumber && f_2(R) = 1 + {2\ep \over 2-\al} {\rho \over \rho_c} = 1 +{1-3\om \over 2(1-2\al -3\om)} \times \\ \nonumber && \left( \sqrt{1 + \ep {16 (1-2\al -3\om) \over (2-\al)(1-3\om)^2} {\ka R \over \rho_c} } - 1 \right)  ~~, \eeqa

\noindent \noindent where the positive branch of the square root was chosen when obtaining $\rho=\rho(R)$ so that $\rho \sim 2\ka R/(1-3\om) $ if $\rho \ll \rho_c$. All expressions collapse to their GR counterparts $f_1(R) \approx R $ and $f_2(R) \approx 1$ when $\rho \ll \rho_c$, as expected.

One may compute the obtained expressions for specific matter contents, namely:

\begin{itemize}
	
	\item Non-relativistic dust ($\om=0$, $\al = 1$):
		
		\beqa R(\rho) &= & {\rho \over 2 \ka} \left( 1 - 2 \ep {\rho \over \rho_c} \right) ~~, \\ \nonumber f_1(R) &=& \ep { \rho_c \over 8\ka } \left( 1 - \sqrt{1 - 16\ep {\ka R \over \rho_c} } \right)   ~~, \\ \nonumber f_2(R) &=& {1 \over 2} \left( 3 - \sqrt{1 - 16\ep  {\ka R \over \rho_c} } \right)  ~~. \eeqa
	
	\item Radiation ($\om=1/3$, $\al = -\om=-1/3$):

		\beqa R(\rho) &=& { 2 \ep \over 7} {\rho^2 \over \ka \rho_c} ~~, \\ \nonumber f_1(R) &=& {R \over 2} ~~, \\ \nonumber f_2(R) &=&  1 + 3 \sqrt{ {2 \ep \over 7} {\ka R \over \rho_c} } ~~.\eeqa
	
	\item Relativistic matter ($\om=1/3$, $\al = 1$):
	
		\beqa R(\rho) &=& -2\ep {\rho^2 \over \ka \rho_c} ~~, \\ \nonumber f_1(R) &=& {R \over 2} ~~, \\ \nonumber f_2(R) &=& 1 - \sqrt{-2\ep {\ka R \over \rho_c}} ~~. \eeqa 
	
	\item Ultrastiff matter ($\om=1$, $\al = 1$):
	
		\beqa R(\rho) &=&   -{\rho \over \ka} \left( 1 + 4 \ep {\rho \over \rho_c} \right) ~~, \\ \nonumber f_1(R) &=& {3 \over 4} R + \ep { \rho_c \over 32\ka } \left( 1 - \sqrt{1 - 16\ep {\ka R \over \rho_c} } \right)  ~~, \\ \nonumber f_2(R) &=& {1 \over 4} \left( 3 + \sqrt{1 - 16\ep {\ka R \over \rho_c} } \right) ~~. \eeqa
	
	\item Scalar field ($\om=1$, $\al = -\om = -1$):
		
		\beqa \label{SFquad} R(\rho) &=&  -{\rho \over \ka} ~~, \\ \nonumber f_1(R) &=& R + {\ep \over 3}{ \ka R^2\over \rho_c} ~~, \\ \nonumber f_2(R) &=& 1 - {2\ep \over 3} {\ka R \over \rho_c}  ~~. \eeqa
	
\end{itemize}

The different forms for $f_i(R)$ obtained for radiation {\it vs.} relativistic matter and for ultrastiff matter {\it vs.} scalar field show the relevance of the choice of the form for $\cl$: although these pairs of matter types are characterized by the same EOS parameter ($\om = 1/3$ and $\om = 1$, respectively), their differing Lagrangian densities (relativistic and ultrastiff matter are assumed to be described by $\cl = -\rho$ (as discussed in Ref. \cite{Lagrangian}), while radiation and a scalar field have $\cl = p$) accounts for the distinct results.

Finally, one notices that a NMC scalar field can lead to a quadratic Friedmann equation, with both a linear NMC and a quadratic curvature term.

%%%%%%%%%%%%%%%%%%%%%%%%%%%%%%%%%%%%%%%%%%%%%%%%%%%%%%%
%%%%%%%%%%%%%%%%%%%%%%%%%%%%%%%%%%%%%%%%%%%%%%%%%%%%%%%

\section{Application: Non-integer power-law form}

The previous section considered a power-law addition to the usual linear term found in the Friedmann equation; another modification of the latter can assume instead that the behaviour of the Friedmann equation is almost linear, so that

\beq \label{NIPower} h(\rho) = \rho \left({\rho \over \rho_c}\right)^\be ~~~~, ~~~~\be \sim 0~~. \eeq

While the previously considered power-law addition may be viewed as arising from a Taylor expansion of a more general modification of the Friedmann equation, the above presents the opposite case of a non-analytical form for $h(\rho)$ (for non-integer $\be$). Regardless of this, both methods developed here still apply, as is shown in the following paragraphs.

\subsection{Perturbative regime}

Since one considers an exponent $\be \sim 0$, the quantity

\beq \label{NIPower2} \de(\rho) \equiv h(\rho) - \rho = \rho \left[ \left({\rho \over \rho_c}\right)^\be - 1 \right] ~~,\eeq

\noindent is a small perturbation, and one may resort to \eq{FriedmannEqPert1} to translate $h(\rho)$ into the corresponding functions $f_1(R)$ and $f_2(R)$. Since the former is linear, its solution (disregarding non-dynamical contributions obeying \eq{FriedmannEqPert1homo1}, as discussed before)  is given by the linear combination of Eqs. (\ref{FriedmannEqPert2Powerf1}) and (\ref{FriedmannEqPert2Powerf2}) with $(n=\be+1, \ep = 1)$ and $(n=1, \ep = -1)$,

\beqa \phi_1(\rho) &=& A_1 {\rho \over \ka} + K_1 {\rho \over \ka} \left({\rho \over \rho_c} \right)^\be ~~, \\ \nonumber \phi_2(\rho) &=&A_2 + K_2 \left({\rho \over \rho_c} \right)^\be ~~, \eeqa

Replacing the above into \eq{FriedmannEqPert1}, one finds that the pairs $(K_1, K_2)$ and $(A_1,A_2)$ still obey \eq{FriedmannEqPert2Power} (with $n=\be+1$, $\ep = 1$ and $n=1$, $\ep = -1$, respectively),

\beqa 1 & =& - \left[ 1 + { 1 + 3\om - 6 (1+\om) \be \over 1-3\om} (\be+1) \right]  K_1 + \\ \nonumber && \left[ 1 + { 1+ 3\om - 6 (1+\om) \be \over 1-3\om} \al \be \right] K_2 ~~,\\ \nonumber  1 & =&  { 2 \over 1-3\om} A_1 - A_2 ~~, \eeqa

\noindent as can be verified explicitly. 

Setting $A_2= K_2=0$, one finds that \eq{NIPower} admits a minimal coupling,

\beqa A_1 &=&{1-3\om \over 2} ~~, \\ \nonumber K_1  &=& - \left[ 1 + { 1 + 3\om - 6 (1+\om) \be \over 1-3\om} (\be+1) \right]^{-1} \\ \nonumber &\approx&  - { 1-3\om \over 2}~~, \eeqa

\noindent so that

\beq f_1(R) \approx R \left(2  - \left[{2\ka R \over (1-3\om) \rho_c} \right]^\be \right) \sim R ~~, \eeq

\noindent and $f_2(R) = 1$.

Conversely, one may resort solely to a NMC, so that $f_1(R) = R $ and

\beqa A_2 &=& -1~~, \\ \nonumber K_2 &=& \left[ 1 + { 1+ 3\om - 6 (1+\om) \be \over 1-3\om} \al \be \right]^{-1} ~~, \eeqa

\noindent thus yielding

\beq f_2(R) \approx \left[{2\ka R \over (1-3\om) \rho_c} \right]^\be ~~. \eeq

Both expressions clearly show the validity of the perturbative regime, since they approach their GR counterpart, $f_2 \sim 1$ when $\be$ vanishes.

\subsection{Relaxed Regime}

Instead of applying the procedure outlined above, one may simply resort to Eqs. (\ref{FriedmannEqPower})-(\ref{f1expPower}) found in the preceding section for a positive addition, $\ep=1$. Replacing the exponent $n \rightarrow 1+ \be$ and disregarding the usual term arising from GR, {\it i.e.} considering only the $\rho_c$ terms, one obtains

\beqa \label{RexpNIPower} R(\rho) &=&  { (1+\be) (1 - 3 \om)  -4 \al \be \over 1+  (1-\al) \be} {\rho \over 2\ka} \left({\rho \over \rho_c}\right)^\be~~, \\ \label{f1expNIPower} f_1(R) &=& \left( 1 +  { 2 \al \be \over (1+\be) (1 - 3 \om)  -4 \al \be } \right) R ~~, \\ \label{FriedmannEqNIPower}   \nonumber f_2(R) &=& { 1 + \be \over 1 + (1-\al) \be} \times \\ &&  \left( { 1 + (1-\al)\be \over ( 1+\be) ( 1-3\om) - 4\al\be} {2\ka R \over \rho_c} \right)^{\be \over 1+\be}~~. \eeqa

Clearly, $\be = 0$ yields the usual GR results. One sees that the deviation from linearity in the Friedmann equation arises from a small correction to the strength of gravity (via the correction to $f_1(R)$) and the new dynamics imprinted by the NMC. Again, both radiation and relativistic matter yield unphysical results, as replacing $\om=1/3$ into the above yields $f_1(R) = R/2$.

%%%%%%%%%%%%%%%%%%%%%%%%%%%%%%%%%%%%%%%%%%%%%%%%%%%%%%%
%%%%%%%%%%%%%%%%%%%%%%%%%%%%%%%%%%%%%%%%%%%%%%%%%%%%%%%

\section{Other modifications of the Friedmann equation}

Another class of modifications of the Friedmann equation relies on the introduction of additional terms on $H$, while keeping the linear energy density term. Indeed, in Ref. \cite{DvaliTurner} the alternative formulation

\beq \label{DT} H^2 - {H^\be \over r_c^{2-\be}} = {\rho \over 6\ka}~~~~,~~~~\be < 2~~, \eeq

\noindent is considered. For a given value of $\be$, inverting the above (even if this can only be attained numerically) yields the form here considered, $H^2 = h(\rho)/6\ka$. One may then apply either of the two methods presented here.

However, the complexity of the ensuing computations (even for a simple linear modification, $\be = 1$, where $H^2=H^2(\rho)$ is easily obtained) implies that an analytic solution is very cumbersome, and perhaps best approached numerically. Since this is not the purpose of this study, the above modification is addressed only in the perturbative regime, as one can advantageously take a further step and insert $H^2 \approx \rho/6\ka$ into \eq{DT}, obtaining

\beq \label{DTperturbative} H^2 = {\rho \over 6\ka} + {1 \over r_c^2} \left( r_cH\right)^\be \approx {\rho \over 6\ka} + {1 \over r_c^2} \left( {r_c^2\rho \over 6\ka} \right)^{\be/2} ~~, \eeq

\noindent showing that the problem collapses into the previously addressed power-law modification, \eq{Power}, with $\ep = 1$, $n = \be/2 $ and $\rho_c = 6\ka /r_c^2$.

%%%%%%%%%%%%%%%%%%%%%%%%%%%%%%%%%%%%%%%%%%%%%%%%%%%%%%%
%%%%%%%%%%%%%%%%%%%%%%%%%%%%%%%%%%%%%%%%%%%%%%%%%%%%%%%

\section{Cosmological Constant from a Nonminimal Coupling}
\label{CCregime}

In this section, one considers the putative relation between a CC and the nonminimally coupled $f(R)$ theory posited by \eq{model}; given the inability of the two methods discussed  in Section \ref{CCapp} to tackle this issue, one approaches it by assuming a weaker condition of exponential expansion of the Universe as a solution of  \eq{FriedmannEq0}. Inserting 

\beq a(t) = a_0 e^{H_0 t} \rightarrow R = 12H_0^2 = 4 \La~~, \eeq

\noindent into the latter, one finds, for the $0-0$ component of \eq{field0},

\beq  \label{field000} \ka f_{01} - f_{02} \rho= 2\left[ \ka F_{01} - (4+3\om)\al F_{02} \rho \right] \La ~~, \eeq

\noindent while the $r-r$ component reads

\beq  \label{field0rr} \ka f_{01} + f_{02} \om \rho = 2 \left[ \ka F_{01} + (4 + 3\om)\om \al F_{02} \rho \right] \La ~~, \eeq

\noindent where one defines $f_{0i} \equiv f_i(4\La) = {\rm constant}$ and $F_{0i} \equiv F_i(4\La) = {\rm constant}$ and uses the covariant conservation of energy-momentum, $\dot{\rho} =- 3H(1+\om) \rho $, which stems from \eq{cov1}, since the scalar curvature is constant. Clearly, the constraint \eq{constraint} cannot be enforced here, since $f_{01}$ and $f_{02}$ are constants, while $\rho $ varies (as argued previously in Ref. \cite{DE1}).

Inspection shows that Eqs. (\ref{field000}) and (\ref{field0rr}) are composed of both constant terms and those linear in the energy density. Equating these one obtains, for $\om \neq -1$,

\beqa  \label{field0setLambdaf1}  f_{01} &=& 2 F_{01} \La ~~, \\ \label{field0setLambdaf2} f_{02} &=& 2 (4+3\om)\al F_{02} \La ~~, \eeqa

\noindent which, again, are not differential equations for $f_1(R) $ and $f_2(R)$, but algebraic ones relating the value of the relevant quantities evaluated at the exponentially expanding phase with a constant scalar curvature.

If one assumes a minimal coupling $f_{02} = f_2(R) = 1$, then \eq{field0setLambdaf2} is ill-defined, since this phase is only obtained if the energy density contribution is negligible. This amounts to considering only \eq{field0setLambdaf1}: a trivial solution is, as expected, given by $f_1(R) = R - 2\La \rightarrow f_{10} = 2\La$ and $F_{01} = 1$ --- although other forms are allowed, such as $f_1(R) = 2\La e^{ (R-4\La)/2\La } \approx R - 2\La$.

The two expressions for $f_1(R)$ presented above are in fact solutions of the differential equation

\beq \label{difeqf1} f_1(R) = 2 f'_1(R) \La~~, \eeq

\noindent and as such hold for the particular scenario of a De Sitter phase, $R = 4\La = {\rm constant}$.

However, one is not restricted to these general solutions: any form for $f_1(R)$ is admissible, even if they do not obey \eq{difeqf1}. If so, \eq{field0setLambdaf1} sets the scale of $\La$ --- as can be seen from the example below, where a power-law form is considered,

\beq \label{f1powerlawCC} f_1(R) = R_1 \left( 1 - {R \over 4\La_1} \right)^n \rightarrow F_1(R) = {n f_1(R) \over R - 4\La_1 }~~. \eeq

\noindent Replacing into \eq{field0setLambdaf1}, one gets

\beq \La = {\La_1 \over 1-{n\over 2}}~~. \eeq

\noindent This, of course, does not shed any light on the CC problem (see Refs. \cite{Weinberg,OBguide}) and why the CC has its observed value, as it merely shifts the question to the value of the parameter $\La_1$.

Also, notice that the strength $R_1$ is not constrained: demanding that $f_1(4\La) = 2\La$, so that the value of the curvature term is identical to its GR value in a De Sitter phase, then

\beq R_1= 2\La \left( {n-2 \over n} \right)^n ~~, \eeq

\noindent and one may rewrite \eq{f1powerlawCC} as

\beq f_1(R) = 2 \La \left( 1 + {R - 4 \La\over 2n\La } \right)^n ~~. \eeq

If one also considers a NMC, then \eq{field0setLambdaf2} has to be considered. Since the curvature term cannot be set to its GR form $f_1(R) = R$ (as this violates \eq{field0setLambdaf1}), this merely adds a layer of complexity to the problem.

However, this can be circumvented if one assumes that the density terms in Eqs. (\ref{field000}) and (\ref{field0rr}) are much larger than the constant contributions, so that \eq{field0setLambdaf1} can be safely neglected; setting $f_1(R) = R$, this yields $\al F_{02} \rho \gg 1$ (as a remark, a power-law expansion $a(t) \sim t^\be$ required the inverse inequality \cite{DE1}). 

A straightforward solution for \eq{field0setLambdaf2} is given by

\beq \label{NMCCClinear} f_2(R) = 1 +  {2 \over (4+3\om)\al -2} {R \over 4\La} ~~.\eeq

\noindent where the integration constant was chosen so that $f_2(0)=1$; notice that the denominator $(4+3\om)\al-2$ is always positive or negative, respectively for either $\al = 1$ or $\al = -\om$ and a positive EOS parameter $\om>0$.

As discussed above, a more complex NMC such as

\beq f_2(R) = \exp \left[{R \over 2(4+3\om) \al \La}\right] ~~,\eeq

\noindent and a power-law form

\beq f_2(R) = \left[ 1 + { R - 4\La \over 2\La \left(4+3 \om\right) n \al } \right]^n ~~, \eeq

\noindent are also suitable.

\section{The Cosmological Constant Problem}

The preceding section shows that a CC may be obtained from a suitable NMC, if the contribution of the time evolving energy density dominates the modified dynamics, embodied by \eq{FriedmannEq0}. This stemmed from the decomposition of the terms in the latter into  constant or evolving with $\rho$, and precluded the possibility that the energy density is also constant.

In this section, one addresses the possibility that a matter species with EOS $ \rho = -p $ dominates the dynamics, so that one has $\om = -1$ and $\al = 1$, regardless of the choice of Lagrangian density (since $\al = 1 $ or $\al = -\om$). Inspection of \eq{cov1} shows that the energy density is then constant, $\dot{\rho} = 0$, regardless of the form for the NMC.

One thus inserts $ - p = \rho  \equiv \rho_{\La}$ into Eqs. (\ref{field000}) and (\ref{field0rr}), which yield

\beq  \label{field0om-1}2 \La = {\ka f_{01} - f_{02} \rho_\La \over \ka F_{01} - F_{02} \rho_\La } = { f_{01} - 2 f_{02} \La_0 \over F_{01} - 2 F_{02} \La_0 } ~~, \eeq

\noindent where one defines $ \La_0 = \rho_\La / 2\ka$. Naturally, setting $f_{01} = 4\La $ and $f_{02}= 1$ yields $\La = \La_0$.

The Cosmological Constant problem lies in the fact that there are approximately $120$ orders of magnitude between the observed and expected value for the CC, $\La \sim \La_0 \times 10^{-120}$, assuming that $\rho_\La$ expresses the energy density of the quantum vacuum \cite{Weinberg}.

Clearly, one can set appropriate forms for $f_1(R)$ and $f_2(R)$ so that \eq{field0om-1} is satisfied. The general implications of considering either a non-linear curvature term or a NMC are outlined below. One begins by assuming a minimal coupling $f_2(R)=1$, so that the above collapses into 

\beq 2\La = { f_{01} - 2 \La_0 \over F_{01} } \rightarrow F_{01} = { f_{01} - 2 \La_0 \over 2\La }  ~~. \eeq

\noindent If one further requires that the non-linear curvature is perturbative, $f_{01} \sim R = 4\La$ but $F_{01} \neq 1$, this becomes

\beq \label{CConlyf1} F_{01} \approx 2 - { \La_0 \over \La } \approx -{ \La_0 \over \La } \sim -10^{120}~~.  \eeq

Conversely, if one only assumes a perturbative NMC, $f_1(R) = R$ and $f_{02} \sim 1$, but $F_{02} \neq 0$, \eq{field0om-1} becomes

\beq  \label{CConlyf2} F_{02} = { f_{02}  \over 2\La } - {1\over 2\La_0} \approx {1 \over 2\La} ~~. \eeq

\noindent This condition is satisfied {\it e.g.} by the NMC in \eq{NMCCClinear}, as can be checked by substituting $\al = -\om= 1$.

One finds that a putative solution of the CC problem using a non-linear curvature term $f_1(R) \neq R$ requires a new dimensionless scale $F_{01} \sim - \La_0/\La$ to reconcile the $120$ order of magnitude difference between $\La$ and $\La_0$, as shown in \eq{CConlyf1}. On the contrary, the observed value $\La$ arises naturally if only a NMC is considered with a characteristic scale $F_{02} \approx 1/(2\La)$.

\section{Discussions and Outlook}

In this work, we have shown that phenomenological modifications of the Friedmann expansion rate equation are related to the fundamental form of the action functional, {\it i.e.} the curvature term $f_1(R)$ and NMC $f_2(R)$.

We have proposed two methods to relate these functions with the modifications of the Friedmann equation: the first method assumes that the latter are perturbative, and expands upon the previous work reported in Ref. \cite{Sotiriou}, while the second relies instead on the condition $F_1 -\al F_2/\ka = {\rm constant}$, which can only be implemented in NMC models.

We have shown that both methods successfully translate a number of specific modifications of the Friedmann equation found in the literature, using both non-trivial functions $f_1(R)$ and $f_2(R)$, or just a non-trivial $f_1(R) \neq R$ ({\it i.e.} $f(R)$ theories) or a non-minimal $f_2(R) \neq 1$.

We have also addressed the possibility of replicating a phase of accelerated expansion of the Universe, by considering the impact of a constant scalar curvature on the modified field equations. We find that the latter is compatible with a nonminimally coupled perfect fluid with any EOS parameter $\om \neq -1$, provided that Eqs. (\ref{field0setLambdaf1}) and (\ref{field0setLambdaf2}) are satisfied.

Finally, we consider the cosmological constant problem, {\it i.e.} how to reconcile the $120$ order of magnitude difference between the observed value $\La$ for the cosmological constant and its expected value $\La_0$, obtained by considering a perfect fluid with EOS $p_\La = -\rho_\La = {\rm constant}$. We find that this difference can be accounted for by either a non-trivial curvature term $f_1(R) \neq R$ or a NMC $f_2(R) \neq 1$ (or a combination of both).

In the first case, this requires the introduction of a dimensionless quantity $F_{01} \approx -\La_0/\La \sim -10^{120}$ --- which merely reframes the cosmological constant problem, shifting it to the question of what is the origin of such a large number.

The use of only a NMC yields a rather interesting result, namely that, by introducing the characteristic scale $ F_{02} \approx 1/2\La$, the ``bare'' cosmological constant $\La_0$ is driven towards its observed value $\La$, regardless of the former: although this mechanism does not account for the value of $\La$, the $120$ orders of magnitude difference from $\La_0$ is effectively removed from the problem.

\acknowledgments

The work of O.B. and J.P. is partially supported by FCT (Funda\c{c}\~ao para a Ci\^encia e a Tecnologia, Portugal) under the project PTDC/FIS/111362/2009.

\end{document}